\documentclass[fp,onecolumn]{jpsj3}
\usepackage{txfonts}

\usepackage{epsfig}
\usepackage{amssymb}
\usepackage{graphicx}
\usepackage{dcolumn}
\usepackage{bm}
\usepackage{color}
\usepackage{tikz}
\newcommand*\circled[1]{\protect\tikz[baseline=(char.base)]{\protect\node[shape=circle,draw,inner sep=0.3pt] (char) {#1};}}

\begin{document}

\title{Conductance Fluctuations in Disordered 2D Topological Insulator Wires: From Quantum Spin-Hall to Ordinary Phases}

\author{Hsiu-Chuan Hsu$^1$, Ioannis Kleftogiannis$^1$, Guang-Yu Guo$^{1,2}$, V\'ictor A. Gopar$^3$}
\inst{$^1$Department of Physics, National Taiwan University, Taipei 10617, Taiwan \\
$^2$Physics Division, National Center for Theoretical Sciences, Hsinchu 30013, Taiwan \\
$^3$Departamento de F\'isica Te\'orica and BIFI, Universidad de Zaragoza, Pedro Cerbuna 12, E-50009, Zaragoza, Spain}

\date{\today}
\abst{
Impurities and defects are ubiquitous in topological insulators (TIs)
and thus understanding the effects of disorder on electronic transport is important.
We calculate the  distribution of the random conductance fluctuations $P(G)$ of  disordered 2D TI wires
modeled by the Bernevig-Hughes-Zhang (BHZ) Hamiltonian with realistic parameters.  As we show,
the  disorder drives the TIs into different  regimes: metal (M), quantum spin-Hall insulator (QSHI), and ordinary insulator (OI).
By varying the disorder strength and Fermi energy,  we calculate analytically and numerically $P(G)$  across the entire
phase diagram. The conductance fluctuations follow the statistics of the
unitary universality class $\beta=2$. At strong disorder and  high energy, however, the size of the fluctutations $\delta G$  reaches
the universal value  of  the orthogonal symmetry class ($\beta=1$).
At the QSHI-M and QSHI-OI crossovers, the interplay between edge and bulk states plays a key role in the statistical properties of the conductance.
}

\maketitle

\section{Introduction}
Topological insulators (TIs) are currently at the forefront of fundamental research
in condensed matter physics and  their singular electronic properties, such as the existence
of dissipationless  helical edge states,  make them   potential candidates for practical uses in
low power consumption and spin-based electronic devices \cite{Qi2011}. Thus, an extensive
literature covering  fundamental and practical  aspects of TIs already exists \cite{Kane, Hasan, Tsung, Bernevig}.
Most of the  studies  have been concentrated on pristine TIs, although the presence of disorder,
such as impurities or lattice imperfections, is unavoidable in real devices.

The helical edge states of TIs are insensitive to weak disorder due to time-reversal (TR) symmetry,
but strong disorder can still affect the  electronic transport properties of TIs,
as was recently demonstrated \cite{Jiang, Li, Imura, Chen}. For instance, the stability of the edge states
in the presence of  disorder has been studied in two-dimensional (2D) TIs and
an unexpected quantum phase, the topological Anderson insulator, was discovered \cite{Jiang, Li}.
Effects of disorder in three-dimensional TIs have been also investigated
within the Aharonov-Bohm oscillations phenomenon \cite{Zhang, Bardarson,Cho, Du}.

Disorder effects in low dimensional electronic devices  have been a fundamental issue since
early studies in the field of mesoscopic physics, particularly  concerning  the problem of
quantum transport\cite{Imry, Beenakker}.  It is known that multiple coherent scattering
of electrons through disordered quantum wires can give a universal character to the transport
in the sense that the statistical properties of the conductance fluctuations depend on
few and general physical parameters of the system, such as TR and spin-rotation symmetries,
whereas microscopic details are irrelevant. One known result on this respect is the constant
value of the standard deviation of the conductance fluctuations in the diffusive regime, the
so-called universal conductance fluctuations (UCF)\cite{Lee}, depending
only on the presence of TR and spin-rotation symmetries: the UCF takes the
value $\sqrt{{8}/{15 \beta}}$, for $\beta=1,2$, and 4. The symmetry class
$\beta=1$ corresponds to systems that preserve both TR and spin-rotation symmetries,
while $\beta=2$ designates the cases with
broken TR symmetry. If TR symmetry is preserved, but spin-rotation symmetry is broken,
we have the symmetry class $\beta=4$\cite{Beenakker}.
Therefore, a statistical study  of the conductance fluctuations is  naturally linked
to the presence of disorder and symmetries play a fundamental role in the analysis.

Moments  of the conductance random fluctuations such as the average and variance have been recently
experimentally and theoretically investigated in TIs \cite{Li, Matsuo, Kandala, Li2014, Zhaoguo, Takagaki, Choe}.
Although these two conductance moments provide important statistical information,
they  might not be sufficient for a full statistical description, especially when
large conductance fluctuations are  present, as frequently happens in experiments.
A complete statistical description of the conductance fluctuations
is provided by the full distribution of the conductance. To the best of our knowledge,
few works have been devoted to the study of the distribution of the conductance fluctuations in TIs\cite{Kobayashi, Xu2012}.
In Ref. \citenum{Kobayashi}, the conductance fluctuations in a quantum network model are
analyzed via the distribution of transmission eigenvalues, while in Ref. \citenum{Xu2012}  the
Bernevig-Hughes-Zhang (BHZ) tight-binding model has been considered  to study numerically the conductance distribution in topological
Anderson insulators. Thus, a thorough theoretical analysis of the conductance fluctuation in 2D TI
is needed.

In this work, we first calculate the conductance distributions of disordered 2D TI wires
modeled by the  BHZ Hamiltonian describing an inverted InAs/GaSb
quantum-well known for its practical advantages over other 2D TI materials,
including the well-developed material control for fabrication and the tunable band structure
by electric fields \cite{Knez, Liu}. We obtain the phase diagrams of the mean and standard deviation
of the conductance, which reveal three different quantum phases or regimes: the metallic (M), the ordinary insulating (OI),
and the quantum spin-Hall insulating (QSHI) phase. Secondly, to get a full understanding of the
statistical properties in each regime and at the crossovers, we perform a theoretical
analysis based on random matrix theory (RMT)\cite{Mellobook, Barbosa}. We find that the conductance fluctuations
for 2D TI wires follow the $\beta=2$ symmetry class in all the phases and crossovers except
in the strongly disordered metallic region, where the  UCF approach  the value of the symmetry class $\beta=1$.

At the M-QSHI crossover regime,
we need to take into account the different extents of localization of the conducting channels
to describe the asymmetric distribution centered around the average conductance $\langle G\rangle=2$.
For simplicity, the unit of the conductance of $e^2/h$ is omitted throughout this paper.
Similarly, as the system is driven from the QSHI to the OI regime, the conductance deviates significantly from $G=2$,
although its fluctuations can be described as in the OI phase. Our analysis reveals that
the edge states play an important role at the crossover regimes
that involve the QSHI phase and the symmetry of the Hamiltonian alone is not enough
to obtain a full understanding of the statistical properties of the conductance.

\begin{figure*}[htbp]
\includegraphics[width=\textwidth,clip=true]{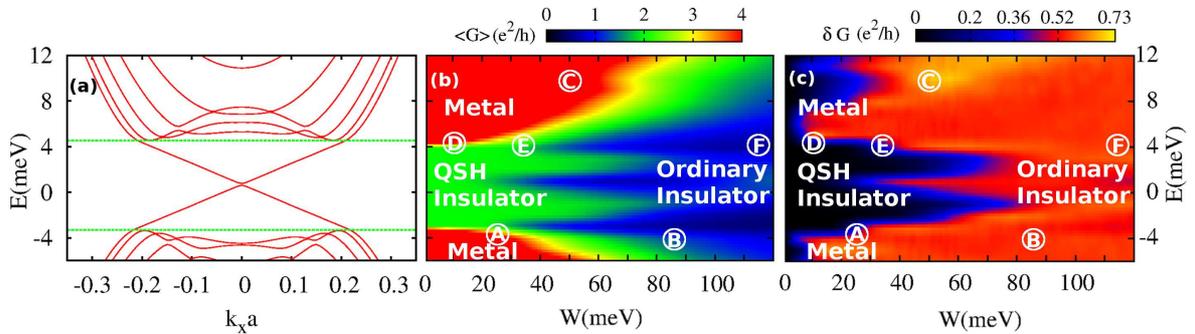}
\caption{(Color online) (a)The band structure of an InAs/GaSb nanowire with $N=60$.
The green dashed lines indicate the energy range where helical edge states occur.
The phase diagrams of (b) the mean conductance $\langle G \rangle$ and (c) the standard deviation $\delta G $
in the $E-W$ plane using 5000 disorder configurations.
Circled letters denote the points where conductance distributions are calculated, as shown in Figs. 2, 4, and 5.}
\label{fig_1}
\end{figure*}

\section{Hamiltonian Model}
We adopt the BHZ model in the tight-binding representation to describe the electronic properties of the clean InAs/GaSb TI wires \cite{Liu,Jiang,Xu}.
The  Hamiltonian $H$ consists of two terms: $H=H_c+ H_d$, where $H_c$ corresponds to a clean (disorder free) TI wire,
and $H_d$ contains a  source of disorder.   The $H_c$ term can be written
in a block diagonal form as
\begin{eqnarray}
	H_c&=&\sum_{i p}V  c^{\dagger}_{i p}c_{i p}
	+ t_{xp} c^{\dagger}_{ip}c_{i+\delta x,p}+t_y c^{\dagger}_{i p}c_{i+\delta y,p}+h.c.\\
	V&=&
	\left(
		\begin{array}{cc}
		 -4D+M-4B & 0    \\
		 0 & -4D-M+4B
		\end{array}
	\right)\\
	t_{xp}&=&
	\left(
		\begin{array}{cc}
 		D+B &  -iAp/2   \\
 		-iAp/2 &  D-B
		\end{array}
	\right)\\
	t_{y}&=&
	\left(
		\begin{array}{cc}
		D+B  & A/2    \\
		 -A/2 &  D-B
		\end{array}
		\right),
\label{Hblock}		
\end{eqnarray}
where $p=\pm$ denotes the pseudo spin, $i=(x,y)$ is the site index and $\delta x(y)$ is the vector to the nearest neighbor along the $x(y)$-direction.
The matrices $V, t_{xp}, t_{y}$ are in the basis of electron and hole states.
The parameters $A,B,D,M$ are material-dependent.
We use the values for those parameters of a InAs/GaSb quantum well derived from $k\cdot p$ method \cite{Liu, Xu}:
$A=18.5$ meV, $B=-165$ meV, $D=-14.5$ meV, $M=-7.8$ meV
with the lattice constant $a=2$ nm.  The Hamiltonian preserves time-reversal symmetry that prevents backscattering of the edge
states \cite{note} as well as the structural and bulk inversion symmetries.
Since we focus on the quantum spin-Hall edge transport, which is protected by the energy gap,
the inversion asymmetries are neglected for simplicity\cite{Liu, Xu}.

A source  of disorder is introduced in the Hamiltonian model in a standard way by considering a random on-site energy:
\begin{eqnarray}
 H_{d}&=&\sum_{ip} \epsilon_{i}c^{\dagger}_{ip}I c_{ip}
 \label{onsitedis}
\end{eqnarray}
where $\epsilon_{i}$ is a random number uniformly distributed in the range $[{-W}/{2},{W}/{2}]$  and $I$ is the $2\times 2$ identity
matrix. Although we have neglected
other sources of disorder that affect the scattering processes such as electron-electron interactions or random fluctuations in
the spin-orbit interaction \cite{Evgeny}, we expect that this
short-range disorder within a single-electron picture captures fundamental physical information of realistic systems. See for instance Ref. \citenum{Du}.

The above BHZ tight-binding Hamiltonian model is solved numerically and the conductance $G$ is calculated by attaching a
perfect lead to each side of the InAs/GaSb wire. Within the Landauer-B\"uttiker formalism \cite{Datta} the conductance is given by
$G=\frac{2 e^2}{h}\sum_i^{n} \tau_i$, where $\tau_i$ are the transmission eigenvalues of $tt^\dagger$, $t$ being
the  $n \times n$ transmission matrix calculated by the recursive Green's function method \cite{Datta, Lewenkopf}
with $n=2N$, where $N$ is the width of the wire in units of the lattice constant.

\section{Results and Discussion}
\subsection{{Phase Diagrams}}

First we show the  band structure of a pristine InAs/GaSb wire in Fig. \ref{fig_1}(a).
Unless stated otherwise, we set the number of sites along the $y$- and $x$-direction to $N = 60$
and $L = 300$, respectively, and assume the transport to be along the $x$-direction.
The edge states occur in the energy window  $[-3.36,4.55]$ meV. The presence of disorder, however,
changes the quantum phase of TIs and some signatures of these phases can be seen in the phase
diagram of the ensemble average conductance $\langle G \rangle$ in the energy-disorder strength ($E-W$) plane
[see Fig. \ref{fig_1} (b)]. From this  phase diagram,  we can
identify the QSHI phase characterized by the average value $\langle G \rangle =2$,
in the  energy range of $[-3.36, 4.55]$ meV (green area) and up to a disorder
strength $W=40$ meV, approximately. If the  energy is increased, within the same range of disorder strength,
the TI wire changes to the metallic regime, whereas if the strength of disorder is increased
within the energy window $[-3.36,  4.55]$meV, the TI wire becomes an ordinary insulator $\left(\langle G \rangle < 1\right)$.

Further useful information  can be obtained by plotting  the  standard deviation of the
conductance $\delta G=\sqrt{(G-\langle G \rangle)^2}$  in the $E$\textendash $W$ plane.
From the $\delta G$ phase diagram shown in Fig. \ref{fig_1} (c),
we observe that for energies in $[-3.36, 4.55]$ meV and until disorder strength $W=40$ meV (black region)
the conductance fluctuations are negligible, revealing  the robustness of  the edge states against disorder in  the
QSHI  regime.  For higher energies,  however, the standard deviation takes the UCF value $\delta G =\sqrt{4/15} \approx 0.52$,
which is  the representative value of the metallic phase for  $\beta=2$.  We notice, however,
that at strong disorder strength and high energy, $\delta G$  reaches the value 0.73
(yellow region in Fig. \ref{fig_1} (c)), which corresponds to the UCF for the symmetry class $\beta=1$.

Once we have identifed the strength of disorder and energy intervals where the different phases of the TI lie,
we proceed to perform a statistical analysis of the conductance fluctuations in each regime.

\subsection{Metal - Ordinary Insulator Crossover}
\begin{figure}
\begin{center}
\includegraphics[width=\columnwidth,clip=true]{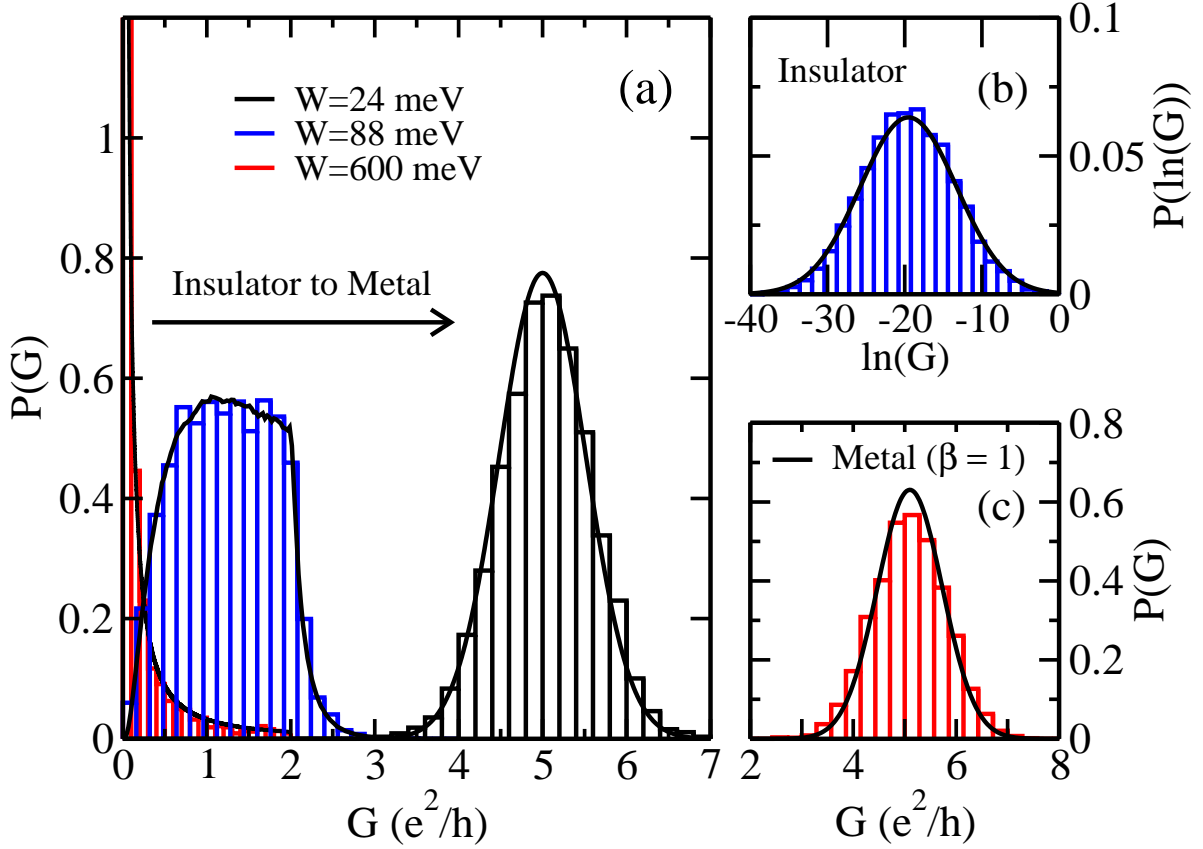}
\end{center}
\caption{(Color online) (a)Conductance distributions at $E=-4$ meV for different disorder strengths $W$ showing the crossover from
M to OI phase. $W=24/88$ meV corresponds to the point \circled{A}/\circled{B} in the phase diagrams in Fig. 1.
The mean conductance $\langle G \rangle=5, 1.2, 0.07$ for $W=24, 88$ and $600$ meV, respectively.
(b) The log-normal distribution  for the OI phase at $E=-4, W=1000$ meV
($\langle G\rangle=5\times10^{-4}, \langle \ln(G)\rangle=-19.5$).
(c) Conductance distribution of the point \circled{C} in the phase diagrams in Fig. 1. The solid line
corresponds to a Gaussian distribution with variance 8/15 (as in the $\beta=1$ symmetry class) and average conductance 5.1.}
\label{fig_2}
\end{figure}
We first study  the evolution of the conductance distribution from  the M  to the OI regimes
as the strength of disorder increases. In Fig. \ref{fig_2} (a), we show $P(G)$ for
three different values of the $W$ at fixed energy $E=-4$ meV.
The histograms are obtained by numerical calculation using $10000$ disorder realizations.
In the M regime, at disorder $W=24$ meV, the conductance distribution follows the expected Gaussian distribution (solid line).
The variance of the Gaussian distribution is given by the UCF: $\sqrt{4/15}$, i.e., $P(G)$ is given by:
\begin{equation}
\label{pofg_metal}
 P(G)=\sqrt{\frac{15}{ 8  \pi}} e^{-\frac{15}{8} (G-\langle G \rangle)^2}
\end{equation}
with $\langle G \rangle=5$ for the case shown in Fig. \ref{fig_2} (a).

We note that the value of the UCF ($\sqrt{{4}/{15}}$) corresponds to the physical systems
with  broken TR symmetry, i.e.,  $\beta=2$ universality class. Indeed,  each block in the
Hamiltonian  breaks TR symmetry and it thus belongs to the $\beta=2$ symmetry, i.e.,
the conductance $G$ is given by the identical contribution of the two Hamiltonian blocks with $\beta=2$ symmetry class.
Moreover, the block diagonal form of the Hamiltonian is a consequence of the inversion symmetry present in our model\cite{Bernevig, Haake}.

As the strength of disorder is increased, the Gaussian distribution becomes wider and the Gaussian landscape is lost. As an example,
in Fig. \ref{fig_2} (a) we show the distribution $P(G)$ in this crossover between the M and OI regime (blue histogram), at disorder
strength $W=88$ meV . In this
case, the conductance fluctuations are large; in particular, for the case shown in Fig. \ref{fig_2} (a),
the relative fluctuations $\delta G/\langle G \rangle$ are five times larger than in the previous metallic phase (black histogram).
The theoretical curve (solid line) that describes the numerical simulations (blue histogram)  has been obtained from the
joint distribution $P_s(\{\tau_i\})$ ($i=1,2,\ldots, n$) of the different transmission
eigenvalues $\tau_i$ together with the Landauer-B\"uttiker expression for the conductance, as follows.
Within the Dorokhov-Mello-Pereyra-Kumar approach \cite{Mellobook}, the joint distribution  for disordered wires supporting $n$ transmission channels, $P(\{\tau_i\})$
where $\tau_i$ are the eigenvalues of the transmission matrix, follows a diffusion equation which
describes the evolution of $P(\{\tau_i\})$ with the system length. For the unitary symmetry class  $\beta=2$, $P(\{\tau_i\})$  is
conveniently written in terms of the variables $x_i=\mathrm{acosh}[1/\sqrt{\tau_i}]$ as \cite{rejaei},
\begin{equation}
\label{pofx}
 p(\left\{x_i \right\})=\frac{1}{Z}\exp{[- H(\left\{ x_i \right\})]},
\end{equation}
where  $Z=\int \Pi_i dx_i \exp[-H(\{ x_i\})]$ and
$ H(x_i)=\sum_{i<j}^n u(x_i,x_j)+\sum_i^n V(x_i)$.

In both the metallic and ordinary insulating regimes, the functions $u(x_i,x_j)$ and $V(x_i)$ can be written as \cite{rejaei, Muttalib}:
\begin{eqnarray}
 u(x_i,x_j)&=& \ln \left| \sinh^2 x_i -\sinh^2 x_j \right|+  \ln \left| x^2_i - x^2_j \right|   ,  \nonumber \\
\label{Vx}
 V(x_i)&=& n s^{-1} x^2_i - \frac{1}{2 }\ln\left| x_i \sinh2x_i\right| ,
\end{eqnarray}
where $s=L/nl$, $l$ being  the mean free path and $n$ the number of conducting channels.
We point out that the ratio $s$ is the only microscopic information that plays a role in the above model and can be determined by the
mean conductance $\langle G \rangle$. For instance, in the metallic limit,  $\langle G \rangle = \xi/L$, where $\xi$ is the localization
length equaling to $nl$, while in the ordinary insulating limit, $\langle G \rangle = e^{-\xi/L}$. In our analysis, $s$ is set to a value
that reproduce $\langle G \rangle$, obtained from the numerical simulations.

Using the joint distribution $P(\{\tau_i\})$ (or equivalently $p(\{x_i\})$) we  calculate the conductance distribution  which is given by
\begin{equation}
\label{pofG_theory}
 P(G)= \big{\langle} \delta (G- 2\sum_i^{n} \tau_i )\big{\rangle} ,
\end{equation}
where the brackets represent the average performed
with the probability density function  $p(\left\{x_i \right\})$,  Eq. (\ref{pofx}). Thus, $P(G)$ in Eq. (\ref{pofG_theory}) can be
calculated by numerical integration.

For the crossover regime shown in Fig. \ref{fig_2} (a), the theoretical distribution $P(G)$ (solid line) is obtained using
the first two channels ($n=2$)
since the conductance values are not larger than 4, indicating that
no more than two different channels from each block Hamiltonian contribute to the conductance,
while  the ratio $s=L/l=5.2$ is found to reproduce the numerical average conductance $\langle G \rangle=1.2$.
Fig. \ref{fig_2} (a)  shows a good agreement between
the theoretical (solid line) and numerical (blue histogram) distributions.

Let us now increase further the strength of disorder to $W=600$ meV. In the presence of  strong disorder, the main contribution to the
conductance comes from a single
transmission eigenvalue and therefore the multivariate distribution $P(\{\tau_i\})$ is simplified to the distribution of a single
variable, $P(\tau)$. In this case, we can write an analytical expression  for the probability density of the conductance $P(G)$ as \cite{Gopar}:
\begin{equation}
\label{pofG_single}
 P(G)=C\frac{\mathrm{acosh^{1/2}(\sqrt{2/G})}}{G^{3/2}(2-G)^{1/4}}\exp \left[(-l/L) \mathrm{acosh}^2\left(\sqrt{2/G}\right)\right] ,
\end{equation}
where $C$ is a normalization constant. As in the previous multichannel case,
the ratio $s= L/l$ plays an important role since all the statistical properties of the
conductance fluctuations are fixed by this ratio. For one channel, $L/l $ can be extracted from the numerical simulations
using $L/l= \langle - \ln (G/2) \rangle $. In  Fig. \ref{fig_2}(a), the conductance distribution (solid line) for $W=600$ meV
according to  Eq. (\ref{pofG_single}) is plotted. A good agreement
between theory (solid line) and the numerical simulations (red histogram) is seen.

If we increase even further the strength of the disorder, the TIs reach the deeply insulating limit and  the
conductance distribution follows a log-normal distribution \cite{rejaei,Muttalib}, as shown for
$W=1000$ meV in Fig. \ref{fig_2} (b). The numerical distribution (histogram) for  $\ln G$ is well described by
$P(\ln G)=1/\sqrt{2\pi\sigma^2}\exp{[-(\ln G - \langle \ln G \rangle)^2/2\sigma^2]}$ with $\sigma^2=2\langle - \ln G \rangle$.
The value of $\langle \ln G \rangle$ has been obtained from the numerical calculations.

\begin{figure}
\begin{center}
\includegraphics[width=0.9\columnwidth,clip=true]{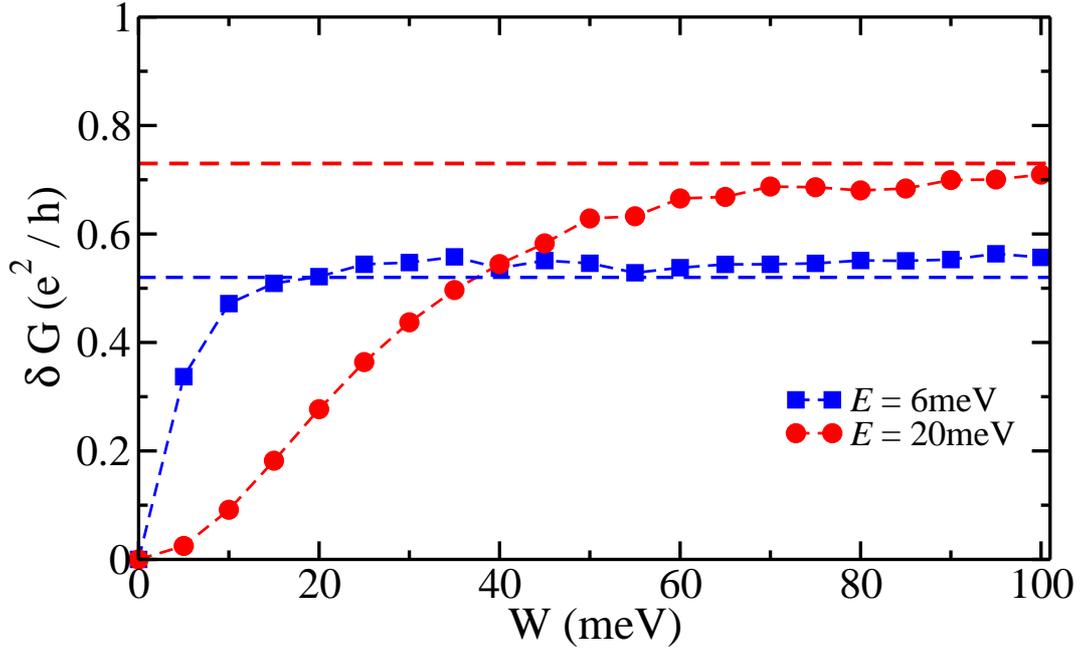}
\end{center}
\caption{(Color online) Standard deviation of the conductance $\delta G$ as a function of the strength of the disorder $W$. At energy $E=6$meV, $\delta G$ is close to the UCF
corresponding to the $\beta=0.52$ symmetry class indicated with a blue dashed-line. However, at $E=20$meV, $\delta G$ approaches the value 0.73 (red dashed-line) that corresponds
to the orthogonal case,  symmetry class $\beta=1$. Notice that we have chosen $E=20$meV which is not within the range of the phase diagram in Fig. 1, in order to clearly show  a
flat behavior of $\delta G$  over a wide range of $W$.}
\label{fig3_re}
\end{figure}

As we have pointed out, the BHZ Hamiltonian model belongs to  the $\beta=2$ symmetry class,   however,  for high energies
($E> 8$ meV) and  disorder strength  $W \sim$50 meV, we notice in the phase diagram  of  Fig. \ref{fig_1} (c),  $\delta G$ approaches the
value $\sqrt{8/15} \approx 0.73$ (yellow region), which corresponds to the UCF for the orthogonal symmetry
($\beta=1$), i.e., the symmetry class  for systems with TR symmetry and no spin orbit coupling (SOC).  Furthermore, we have obtained the distribution of the conductance fluctuations
at the point $\circled{C}$ in Fig. \ref{fig_1} (c) which, as shown in Fig. \ref{fig_2}(c),  follows a Gaussian distribution:  the solid line corresponds to a Gaussian
distribution with variance ${8/15}$, i.e., the variance value of the UCF for the orthogonal symmetry class.
Additionally, to show the robustness of this result and contrast it  with the conductance fluctuations for $\beta=2$,   in Fig. \ref{fig3_re} we plot  $\delta G$  as a function of the
disorder strength $W$ at two different energies: $E=6$meV (blue squares) where the conductance standard deviation  approaches the value 0.52, i.e.,   the $\beta=2$ symmetry
class  and  at  $E=20$meV (red dots),  where $\delta G$ is close to 0.73 that corresponds to the symmetry $\beta=1$.
This essentially reveals that effects of the SOC (off-diagonal terms in the Hamiltonian $H_c$)
are negligible  at high energies and strong disorder. \cite{simulations_TRS, Kaneko}

\begin{figure}
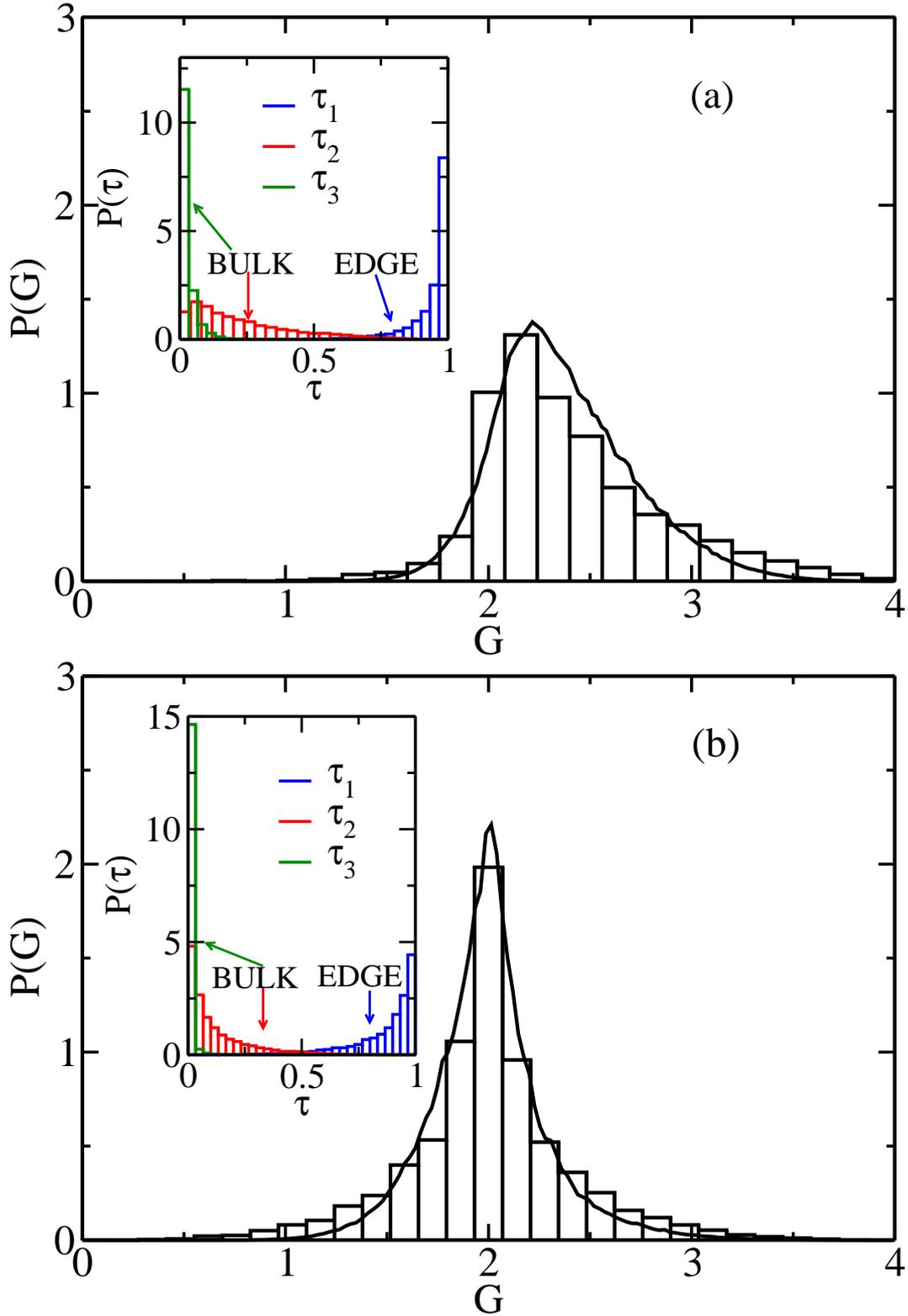

\begin{center}
\includegraphics[width=0.9\columnwidth,clip=true]{fig4a.eps}
\includegraphics[width=0.9\columnwidth,clip=true]{fig4b.eps}
\end{center}
\caption{(Color online) The conductance distribution with $\langle G\rangle=2.39$ at the M-QSHI crossover at $E=4.55$ meV and $W=12$ meV
for two different system lengths (a) L=300, corresponding to \circled{D} in the phase diagrams in Fig. 1 and (b) L=600. The solid line is the theoretical distribution. The
probability distributions of the three largest transmission eigenvalues
$\tau_1$, $\tau_2$, $\tau_3$
are shown in the insets. For the distribution in  (b), $\langle G\rangle=1.99$ and  $s_2=19s_1$, $s_3=44s_1$ with $s_1=0.34$. }
\label{fig4_re}
\end{figure}

\subsection{Metal - Quantum Spin-Hall Crossover }
In the M phase
the conductance fluctuations follow a Gaussian distribution, as shown above. On the contrary, in the QSHI
phase, there is no conductance fluctuations due to the robustness of  the helical edge states against disorder. Thus,
it is of interest to study the conductance fluctuations at the crossover between these two phases,  when both edge and bulk
states contribute to the conductance.

In our numerical simulations, we have chosen the point at the crossover: $W=12, E=4.55$ meV (point \circled{D} in
Fig. \ref{fig_1}(b) and (c)).
At this M-QSHI crossover point, we have analyzed the fluctuations  of the first three largest different eigenvalues, represented by
$\tau_1$, $\tau_2$ and $\tau_3$.  We have verified numerically that
these three channels give the main contribution to total conductance $G$ i.e., the contribution of higher transmission channels is negligible. In
the inset of Fig. \ref{fig4_re}(a) and \ref{fig4_re}(b), we show the distribution of each
transmission eigenvalue for two different TI wire lengths ($L=300$ and 600, respectively).
We notice that the first transmission eigenvalue $\tau_1$, which is associated to the helical edge state,  is highly transmitted,
whereas the second and third transmission eigenvalues   associated to the bulk states  are poorly transmitted.
Hence, the distribution of the first eigenvalue peaks at $\tau_1=1$, while the distribution of
$\tau_2$ and $\tau_3$, respectively,  peaks at zero. This suggests that  helical edge states are weakly localized,
whereas bulk states are strongly localized. Therefore,  in order to describe
the conductance fluctuations within our theory, we need to introduce the information about the different strengths of localization of the edge
and bulk states.

As we have pointed out above, the strength of the electron localization in our  model  is characterized by the
parameter $s=L/l$ (see Eqs. (\ref{Vx}) and (\ref{pofG_single})), which is determined  by $\langle \ln G \rangle$
in the case of one transmission eigenvalue. Thus, in order to consider  different localization strengths,
we propose  introducing  a different $s_i$ value for each transmission eigenvalue, which is extracted from
the numerical calculations. For instance,  for the distribution $P(G)$  shown in Fig. \ref{fig4_re} (a), we
found  numerically that $\langle -\ln \tau_2 \rangle$ is  25 times larger than  $\langle -\ln \tau_1 \rangle$ , while
$\langle -\ln \tau_3 \rangle$ is  60 times larger than $\langle -\ln \tau_1 \rangle$.
Thus, we set $s_2=25 s_1$ and  $s_3=60 s_1$ with $s_1=0.24$ fixed to reproduce the average
conductance $\langle \ln G \rangle$, or equivalently $\langle G \rangle$. As an additional example,
we show $P(G)$ for a longer  TI wire ($L=600$) in Fig. \ref{fig4_re}(b). Thus, the proposed method to calculate $P(G)$ needs
a detailed analysis of the contribution of the different transmission eigenvalues, in contrast to the standard analysis where the knowledge of
the conductance average, or the parameter $s$, is sufficient  to obtain $P(G)$.  Interestingly,
distributions with similar landscapes at the M-QSHI crossover of a Z2 network model
were reported in Ref. [\citenum{Kobayashi}],  where it was also found that the first largest transmission eigenvalues
determine the conductance distributions. In Ref. [\citenum{Kobayashi}], however, the network model belongs to
the symplectic class ($\beta=4$), whereas our case is for the unitary class $\beta=2$.
We thus conclude that the landscape of  the conductance distribution is strongly determined by
the different localization, and therefore different contribution to the conductance, of the  transmission channels
rather than the symmetry class.

\subsection{Ordinary Insulator - Quantum Spin-Hall Crossover}

\begin{figure}
\begin{center}
\includegraphics[width=0.8\columnwidth,clip=true]{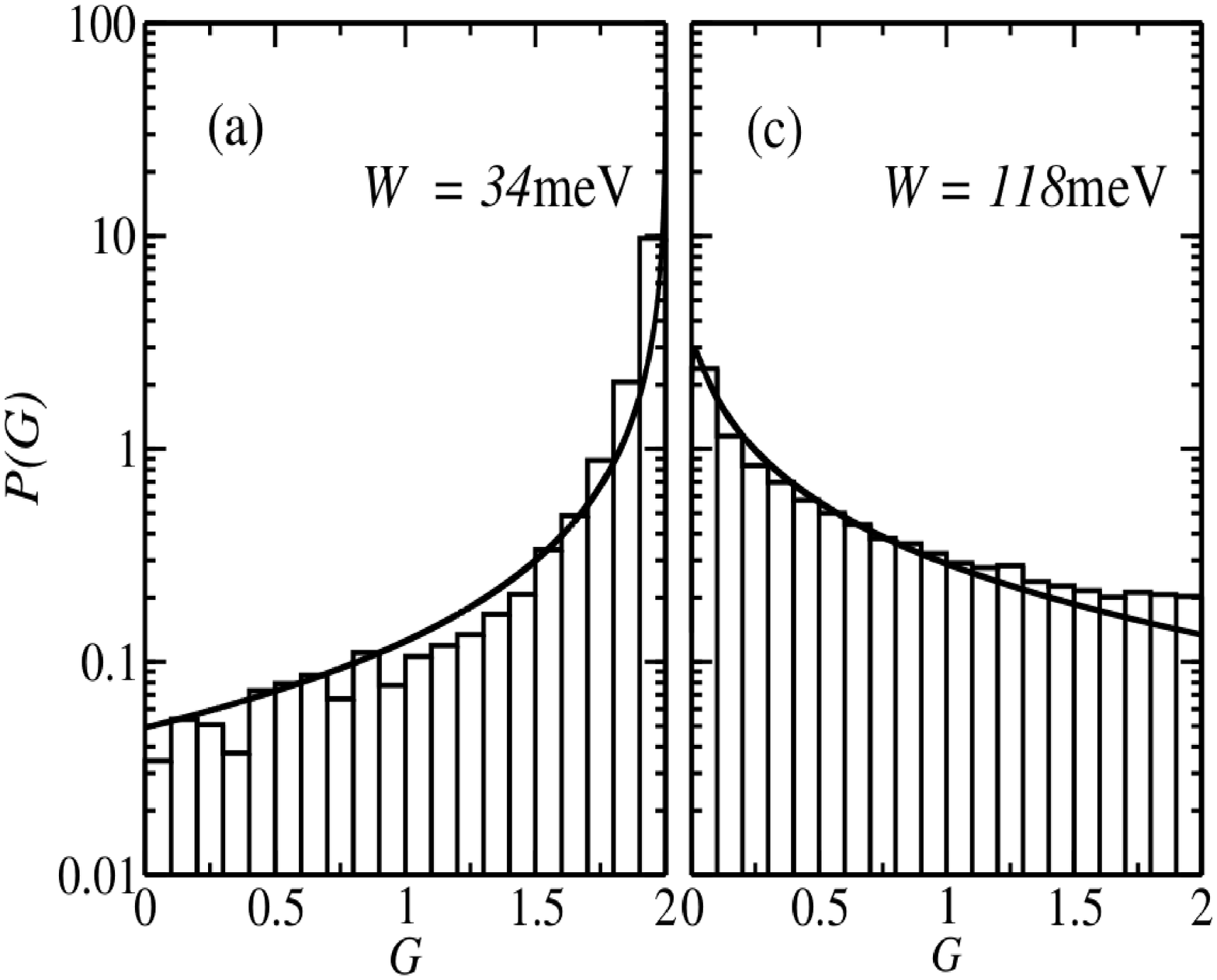}
\includegraphics[width=0.8\columnwidth,clip=true]{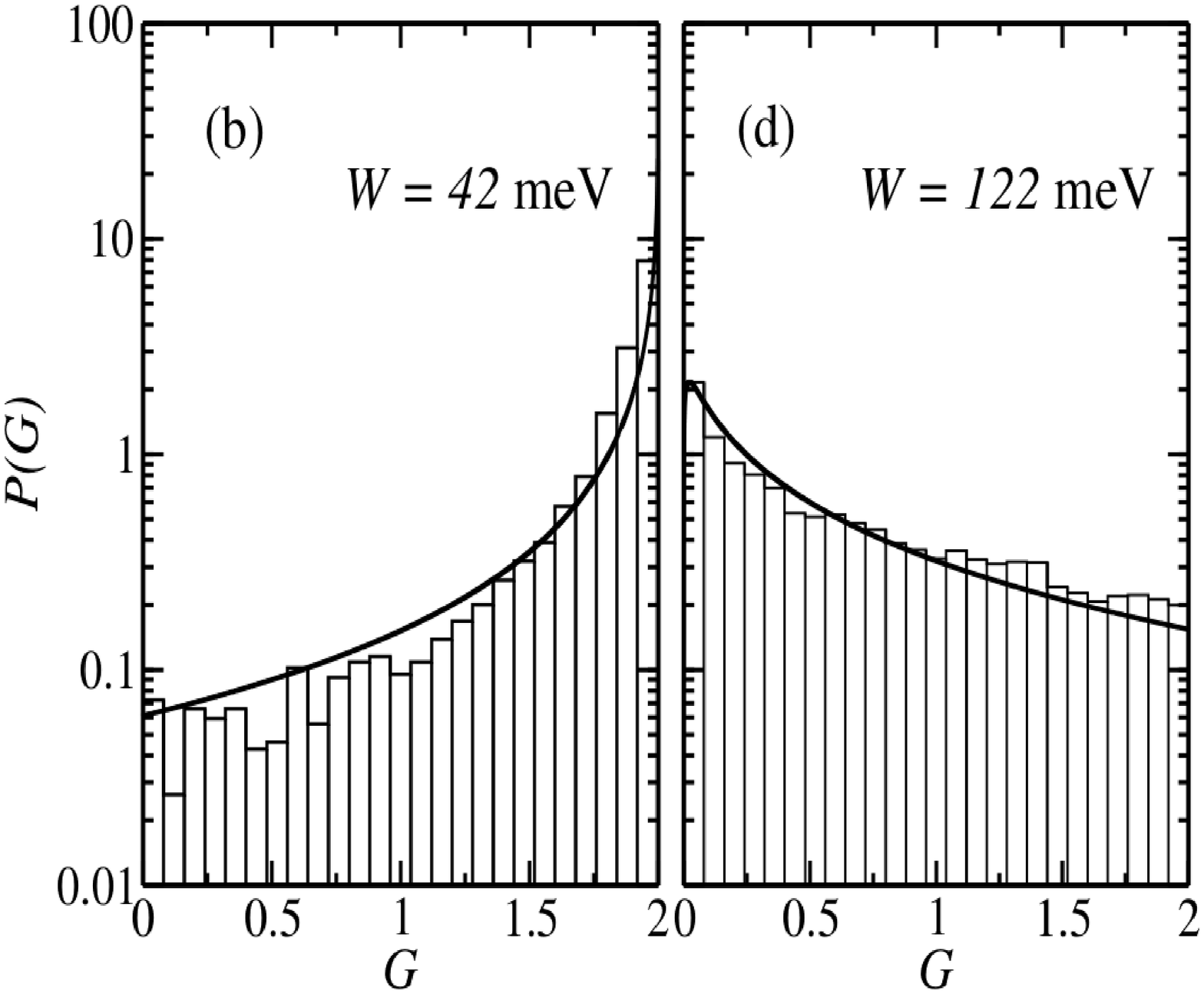}
\end{center}
\caption{Conductance distributions at the QSHI-OI crossover on the QSHI side  [panels (a) and (b)]  and on the OI side
[panels (c) and (d)]. The distributions in panels (a) and (c) correspond to the points \circled{E} and \circled{F} in the phase diagram in Fig. 1, at  $E=4$ meV. For the panels (b) and (d)
$E=3.8$ meV. Histograms and solid lines
curves correspond to the numerical and theoretical distributions: Eq. (\ref{pofG_single})  with $s=2.03$ (a),  $s=1.86$ (b) and Eq. (\ref{pofG_single_iqsh}) with $s'=3.6$ (c),
$s'=3.22$ (d).}
 \label{fig5_re}
\end{figure}

As we have  shown above, helical edge states are robust against disorder. However, at sufficiently strong  disorder,
edge states eventually become localized, leading the QSHI phase to an OI phase.
At this crossover regime, but on  the QSHI side, the conductance distribution has a large peak at $G=2$,
as shown in  Figs. \ref{fig5_re} (a) and \ref{fig5_re} (b) , because the edge states have not yet been fully localized.
In contrast, on  the OI side, the distribution shows a
 peak at $G=0$ [see Figs. \ref{fig5_re} (c) and  \ref{fig5_re} (d)], indicating  that the edge states become localized.

To study the fluctuations on the QSHI side, we assume that the conductance fluctuations come from the edge states
that start to penetrate into the 2D wire and whose conductance fluctuations can be described by Eq. (\ref{pofG_single}), as in the OI phase. On the QSHI side, however,
the maximum conductance of the edge states  is 2; we thus only need to make the change of variable $G \to 2-G$ in
Eq. (\ref{pofG_single}) to obtain the distribution of the conductance fluctuations:
\begin{equation}
\label{pofG_single_iqsh}
 P(G)=C\frac{\mathrm{\sqrt{\mathrm{acosh}\sqrt{\frac{2}{2-G}}}}}{(2-G)^{3/2}G^{1/4}}\exp \left[-s'^{-1} \mathrm{acosh}^2\sqrt{\frac{2}{2-G}}\right] ,
\end{equation}
where $C$ is a normalization constant and $s'$ is determined by the average  $s'=\langle \ln [(2-G)/2]\rangle$.
Fig. \ref{fig5_re} (a) and \ref{fig5_re} (b) show $P(G)$ (solid line) given by Eq. (\ref{pofG_single_iqsh}) which agrees well with  numerical simulations (histograms).

Regarding the OI to QSHI crossover, but on the OI side,
the perfect conducting edge states become localized and the TI becomes an ordinary insulator. Thus,
only a single transmission eigenvalue is relevant and the distribution of
conductances is described by Eq. (\ref{pofG_single}). Figs. \ref{fig5_re} (c) and \ref{fig5_re} (d) show  both
the numerical (histogram) and theoretical (solid line) distribution for $P(G)$.

\section{Summary and Conclusions}

Topological insulators are not free from impurities and/or lattice defects and therefore a full understanding
of the effects of the disorder on the electronic transport is of interest from practical and fundamental points of view.
We have studied the conductance fluctuations of disordered 2D TI wires
modeled by the  BHZ Hamiltonian with experimentally achievable parameters. We have been gone beyond the first moments of the conductance fluctuations
and calculate analytically and numerically the  complete distribution of the conductance.

The phase diagrams
of the mean and the standard deviation of the conductance, which have been numerically obtained, show
three different quantum phases or regimes: metal, quantum spin-Hall insulator, and ordinary insulator.
We have obtained  the conductance distribution in each regime as well as in their  crossovers  within a framework of random matrix theory.
The conductance fluctuations follow the statistics  of the unitary class $\beta=2$  as a consequence of the block diagonal structure
of the Hamiltonian, in which each block belongs to the unitary class $\beta= 2$.  At strong disorder and high energies, we have found that  the
conductance fluctuations  $\delta G$ approach the value of  the orthogonal universality class $\beta=1$. We interpretate this result as a consequence of the fact that  at high energies,
the band structure of the TIs resembles that one of a normal metal and therefore the value of $\delta G$  corresponds to case of the orthogonal universality class, as long as the
disorder is sufficiently strong to wash out the SOC effects.
We have actually  numerically  verified that  the  value of  conductance fluctuations ($\beta=1$) is  not affected   by the strength of the SOC.

Furthermore, our results reveal that at M-QSHI and OI-QSHI crossover regimes, the presence of the helical edges states play a crucial role in the conductance
statistics.  At those crossover regimes,
both bulk and helical edge states contribute to the conductance,
albeit with different extent of  localization.
Consequently, the conductance distributions change drastically from one quantum phase to
another one due to the different degree of localization of the helical edge and bulk states,
as well as the number of relevant transmission channels.

We  believe that the extensive analysis of the conductance fluctuations presented
here gives a complete picture of the statistics of the
conductance fluctuations and offers deeper insights
into the quantum transport in disordered TIs.

\begin{acknowledgment}
H.-C. H. would like to thank Dong-Hui Xu for helpful discussions. V. A. G thanks R. A. Molina for helpful suggestions and discussions.
V. A. G acknowledges support from MINECO (Spain) under the Project number FIS2015-65078-C2-2-P and Subprograma Estatal de Movilidad 2013-2016 under
the Project number PRX16/00166. He also thanks The Physics Department and the Center for Theoretical Sciences of the National Taiwan University,
as well as the Physics Department of Queens College, The City University of New York for the hospitality.
H.-C. H., I. K. and G.-Y. G. acknowledge support from Academia Sinica, National Center for Theoretical
Sciences and the Ministry of Science and Technology of The R.O.C.
The authors gratefully acknowledge the resources from the supercomputer ``Terminus", technical expertise
and assistance provided by the Institute for Biocomputation and Physics of Complex Systems (BIFI) -
Universidad de Zaragoza as well as from the National Center of High-performance Computing of Taiwan.
\end{acknowledgment}


\begin{thebibliography}{99}
\bibitem{note} We point out that, contrary to the model studied here,  in the so-called topological Anderson
insulator introduced in \citenum{Li}, there is
no inversion of the bands (the mass term was fixed to a positive value), i.e., edge states are absent, in the clean limit. The
existence of edge states was shown at intermediate
values of the disorder strength.
\bibitem{simulations_TRS}
We have  verifed numerically that statement: At $E=20$meV we have calculated the standard deviation $\delta G$ at different values of the SOC
($A=$18.5, 12, and 6 meV) and noticed  that $\delta G$ is not affected, i.e., $\delta G \sim 0.73$ in all cases. In contrast, at low
energies, for instance  at $E=6$meV the fluctuations of the conductance depend on the value of the SOC.
\end{thebibliography}
\end{document}